\documentclass[letterpaper,aps,pra,twocolumn,notitlepage]{revtex4}

\usepackage[utf8]{inputenc}
\usepackage{amsmath}
\usepackage{amsfonts}
\usepackage{amssymb}
\usepackage{graphicx}
\usepackage{braket}
\usepackage{lipsum}
\usepackage[margin=2cm]{geometry}
\newcommand{\hc}{\text{H.c.}}
\usepackage{fullpage}
\usepackage{appendix}
\usepackage{comment}
\usepackage{color}
\usepackage{hyperref}

\newcommand{\dg}{^{\dagger}}

\begin{document}
\title{Broadband pseudothermal states with tunable spectral coherence generated via nonlinear optics}
\author{Nicol\'as Quesada}
\affiliation{Perimeter Institute for Theoretical Physics, Waterloo, Ontario, Canada, N2L 2Y5}
\author{Agata M. Bra\'nczyk}
\email{abranczyk@pitp.ca}
\affiliation{Perimeter Institute for Theoretical Physics, Waterloo, Ontario, Canada, N2L 2Y5}

\begin{abstract}
It is well known that the reduced state of a two-mode squeezed vacuum state is a thermal state---i.e. a state whose photon-number statistics obey a geometric distribution. More exotic \emph{broadband} states can be realized as the reduced state of two spectrally-entangled beams generated using nonlinear optics.  We show that these broadband ``pseudothermal'' states are tensor products of states in spectral Schmidt modes, whose photon-number statistics obey a geometric distribution. We study the spectral and temporal coherence properties of these states and show that their spectral coherence  can be tuned---from perfect  coherence to complete  incoherence---by adjusting the pump  spectral width. In the limit of a cw pump, these states are tensor products of true thermal states, but with different temperatures at each frequency. This could be an interesting state of light for investigating the interplay between spectral, temporal, and photon-number coherences. 
\end{abstract}

\maketitle

\section{Introduction}
Thermal states are of fundamental and practical interest. Although they are diagonal in the photon-number and coherent-state bases, they can behave non-classically. They can be used for generating non-classical  states \cite{Zavatta2007,Tahira2009}, or for mediating entanglement between quantum systems \cite{Kim2002a}. They can also be used  for quantum information protocols such as continuous-variable quantum key distribution \cite{Weedbrook2012}, and improving the efficiency of quantum state tomography \cite{Harder2014}. 

In quantum information, one often deals with single-mode thermal states. There, the relevant property is the photon-number statistics---the thermal state density matrix is diagonal  with a geometric probability distribution. 

In quantum optics, when considering many radiation modes, the temperature $T$ takes a more central role as it determines the light's spectral radiance according to Planck's law. For multimode thermal light, one can  speak about its spatial, spectral, temporal, and momentum coherence. 

Incoherence in these degrees of freedom can be useful. Spatially incoherent light has been used for ghost imaging \cite{Gatti2004a,Gatti2004,Valencia2005,Zhang2005b,Agafonov2009,Chen2010,Moreau2017}, sub-wavelength lithography \cite{Cao2010}, and  improving diffraction pattern visibility \cite{Lu2008} and  spatial resolution \cite{Sprigg2016}. Broadband spectrally incoherent light has been used for  resolution-enhanced optical coherence tomography  \cite{Lajunen2009}, optical guiding of microscopic particles \cite{Lopez-Mariscal2010}, and noisy-light spectroscopy  \cite{Ulness2003,Turner2013}. 

Various methods exist for generating incoherent light. Spatially incoherent, pseudothermal light can be generated by sending a cw laser   through a rotating ground-glass disc \cite{Li2005,Iskhakov2011,Zhu2012a}.   Broadband, spectrally incoherent light can be generated by a thermal source---e.g. a hollow cathode lamp  \cite{Zhang2005b}. Other approaches use amplified spontaneous emission from quantum dots, \cite{Jechow2013,Hartmann2015},  warm atomic vapour \cite{Mika2018}, and modified dye lasers \cite{Schulz2005}. Here, we focus on  light generated in one arm of a broadband twin-beam state, such as that generated via spontaneous parametric downconversion (SPDC) \cite{Christ2011,Spasibko2017} or spontaneous four wave mixing (SFWM) \cite{Vernon2017}. Spectral entanglement between the beams makes each individual beam spectrally incoherent.

Photon statistics and coherence can be studied by measuring correlation functions. Various groups measured such functions on light generated via nonlinear optics. These include measurements of time-resolved temporal correlation functions in cw-pumped SPDC \cite{Blauensteiner2009}, time-averaged temporal correlation functions in two-mode SPDC \cite{Christ2011,eckstein2011highly},  frequency cross- and auto-correlation functions in two-mode SPDC \cite{Spasibko2012}, multi-photon statistics of single-mode SPDC \cite{Wakui2014}, and frequency-resolved spectral correlation functions and multi-photon statistics in harmonic generation \cite{Spasibko2017}. Theory has also been done on photon-number statistics \cite{Huang1989} and spatial correlation functions \cite{Gatti2004,Gatti2004a} of downconverted light. 

But to the best of our knowledge, no one has written down the reduced density operator for one arm of an arbitrary spectrally-entangled twin-beam state, nor has anyone computed its time-resolved temporal and frequency-resolved spectral correlation functions. 

In this paper, we show that the density operator describing one arm of a twin-beam state with arbitrary spectral entanglement can be decomposed into a tensor product of states prepared in spectral Schmidt modes, each with geometric photon-number statistics. We therefore refer to the state as a broadband ``pseudothermal'' (BPT) state. 

We then write down Heisenberg-picture operators for the Schmidt modes and compute the BPT state's frequency-resolved spectral and time-resolved first- and second-order correlation functions. These functions reveal the light's spectral and temporal coherence, as well as its intensity-intensity correlations. We find that the spectral coherence of BPT states can be tuned---from perfect  coherence to full  incoherence---by adjusting the pump spectral width. This is consistent with recent experiments \cite{eckstein2011highly}. (The spectral coherence can also tuned by changing the spectral phase of the pump, e.g. by applying a chirp \cite{ansari2018tomography}, although we do not consider this  here).   From these correlation functions, one can identify interesting regimes not yet explored experimentally, such as partial spectral coherence.

In summary, the main contributions of this paper are three-fold. First, we derive the reduced density operator of one beam of a twin-beam state with arbitrary spectral entanglement, in terms of its Schmidt modes. This expression has conceptual value and provides intuition about the nature of the light in each beam. Second, we derive time- and frequency-resolved correlation functions for one beam of a twin-beam state in terms of its Schmidt modes, and explore the relationship between inter-beam spectral entanglement and single-beam temporal coherence.  Third, we make clear the connection between these correlation functions and results from classical coherence theory. Our results should therefore be useful for both experimental and theoretical studies of partially-spectrally-coherent light. 

\section{Broadband two-mode squeezed vacuum state}

We consider broadband light, generated by e.g. SPDC or SFWM, emitted into two orthogonal modes. In a one dimensional propagation geometry where the longitudinal wave vector is specified by the frequency, and assuming that the pump beam remains un-depleted,  the twin-beam state is:
\begin{align}
 \ket{\psi} ={}&\mathcal{\hat U}_{\text{SQ}} \ket{\text{vac}}\,;\\
 \mathcal{\hat U}_{\text{SQ}} =  {}& e^{ \left(\iint \mathrm d \omega_a  \mathrm d \omega_b\, J(\omega_a, \omega_b) \hat{a}^\dagger(\omega_a) \hat{b}^\dagger(\omega_b) - \hc \right) },
\end{align}
where $J(\omega_a, \omega_b)$ is known as the \emph{joint spectral amplitude} (JSA) of the generated beams \cite{Grice1997,Quesada2014,Quesada2015} and $\mathcal{\hat U}_{\text{SQ}}$ is the broadband two-mode squeezing operator. The operators $\hat{a}(\omega) $ and $\hat{b}(\omega) $ are single-frequency annihilation operators that satisfy the commutation relations $[\hat{a}(\omega),\hat{a}\dg(\omega')]=[\hat{b}(\omega),\hat{b}\dg(\omega')]=\delta(\omega-\omega')$ (all other commutators are zero). The JSA depends on the properties of the pump field(s) and the nonlinear material. For purposes of this paper, we leave it quite general. 

To simplify calculations, the JSA can be decomposed as $J(\omega_a, \omega_b)=\sum_{k} r_k \phi_k(\omega_a)\varphi_k(\omega_b)$, in what is known as the Schmidt decomposition.  The twin-beam state can then be written as \cite{Christ2011}:
\begin{align}\label{eq:schmidt}
 \ket{\psi} ={}&e^{\sum_k r_k \hat{A}_k^\dagger \hat{B}_k^\dagger - \hc }\ket{\mathrm{vac}}\,,  
\end{align}
where the operators
\begin{subequations}\label{eq:bigops}
\begin{align}
    \hat{A}_k ={}& \int \mathrm d \omega_a \phi_k^*(\omega_a) \hat{a}(\omega_a)\,, \\
    \hat{B}_k ={}& \int \mathrm d \omega_b \varphi_k^*(\omega_b) \hat{b}(\omega_b)\,,
\end{align}
    \end{subequations}
are broadband annihilation operators that satisfy the commutation relations $[\hat{A}_k,\hat{A}\dg_{k'}]=[\hat{B}_k,\hat{B}\dg_{k'}]=\delta_{k,k'}$ (all other commutators are zero), and $\phi_k$ and $\varphi_k$ are known as Schmidt modes; these functions satisfy completeness and orthogonality relations \cite{quesada2018b}. 

Analytical forms for the Schmidt decomposition  are known only for two-dimensional Gaussian functions \cite{mehler1866ueber}. For more general functions, one can use approximate numerical methods, e.g., computing the singular value decomposition of a truncated, discretized JSA. The impact of such approximations in SPDC source characterization was recently discussed in \cite{Graffitti2018b}. 

One can also invert the relations in Eqs. (\ref{eq:bigops}) and find
\begin{subequations}\label{eq:invert}
\begin{align}
\hat{a}(\omega)={}&\sum_k \phi_k(\omega)\hat{A}_k\,,\\
\hat{b}(\omega)={}&\sum_k \varphi_k(\omega)\hat{B}_k\,.
\end{align}
\end{subequations}
The twin-beam state in Eq. (\ref{eq:schmidt}) can be rewritten as:
\begin{align}\label{eq:schmidt2}
 \ket{\psi}   ={} \bigotimes_k \ket{r_k}\,,
\end{align}
where
\begin{align}\label{eq:sq}
 \ket{r_k}=e^{r_k \hat{A}_k^\dagger \hat{B}_k^\dagger - \hc }\ket{\mathrm{vac}}
\end{align}
is a two-mode squeezed vacuum (TMSV) state prepared in two Schmidt modes $\phi_k$ and $\varphi_k$, with \emph{squeezing parameter} $r_k \geq 0$ . Note that here $J(\omega_a, \omega_b)$ is not necessarily normalized and thus $\sum_k r^2_k$ does not necessarily equal 1. 
Each TMSV state can be represented in the  number basis 
\begin{align}
 \ket{r_k} ={}&\sum_{n_k=0}^{\infty}\frac{(\tanh r_k)^{n_k}}{\cosh(r_k)}\ket{n_k}_{\phi_k}\ket{n_k}_{\varphi_k}\,.
\end{align}

In the Heisenberg picture, the operators $\hat{A}_k$ and $\hat{B}_k$ transform as:
\begin{subequations}\label{HeisenbergEq}
	\begin{align}
	\begin{split}
	\hat{A}_k \to{}&  \mathcal{\hat U}_{\text{SQ}}^\dagger \hat{A}_k  \mathcal{\hat U}_{\text{SQ}} \\
	={}& \cosh(r_k) \hat{A}_k+ \sinh(r_k) \hat{B}_k^\dagger\,, 
	\end{split}\\
	\begin{split}
	\hat{B}_k \to{}&  \mathcal{\hat U}_{\text{SQ}}^\dagger \hat{B}_k  \mathcal{\hat U}_{\text{SQ}}\\
	={}& \cosh(r_k) \hat{B}_k+ \sinh(r_k) \hat{A}_k^\dagger\,.
	\end{split}
	\end{align}
\end{subequations}
Later, we will use the relations in Eqs. (\ref{eq:invert}) and the transformation in Eqs. (\ref{HeisenbergEq}) to compute correlation functions.

\section{Broadband pseudothermal states}
We are interested in the quantum state of the individual beams, we thus compute the reduced density matrices for modes $a$ and $b$
 by tracing out the other mode. The density operator for  mode $a$ is
\begin{align}\label{eq:rhoa}
\rho_a={}&\mathrm{Tr}_b\big[ \ket{\psi} \bra{\psi}\big]=\bigotimes_k \rho_{\phi_k}\,,
\end{align}
where
\begin{align}\label{gibbs0}
\rho_{\phi_k}={}&\mathrm{Tr}_{\varphi_k}\big[  \ket{r_k} \bra{r_k}\big]\\\label{eq:phi_k}
={}&\sum_{n_k=0}^{\infty}P_{n_k}\ket{n_k}_{\phi_k}\bra{n_k}_{\phi_k}\,,
\end{align}
is a state prepared in a single Schmidt mode $\phi_k$. The states $\ket{n_k}_{\phi_k}=(n_k!)^{-1/2}(    \hat{A}\dg_k)^n\ket{0}$ are broadband Fock states, and  are distributed according to 
\begin{align}\label{gibbs1}
P_{n_k}=\frac{1}{1+\bar{n}_k}\left(\frac{\bar n_k}{1+\bar n_k}\right)^{n_k}\,,
\end{align}
where 
\begin{align}\label{eq:nk}
\bar{n}_{k}=\sinh^2(r_k)\,.
\end{align} 
The state $\rho_{\phi_k}$ is like a single-mode thermal state in the sense that it is diagonal in the photon-number basis and the probability distribution is geometric. But since the mode is not at a well-defined frequency, it doesn't make sense to talk about an associated temperature $T$.

Similarly, the beam in mode $b$ has the state $\rho_b=\bigotimes_k \rho_{\varphi_k}$ where $\rho_{\varphi_k}=\sum_{n_k=0}^{\infty}P_{n_k}\ket{n_k}_{\varphi_k}\bra{n_k}_{\varphi_k}$. 
While both $\rho_a$ and $\rho_b$ obey the same statistics given by $P_{n_k}$, the states will  have different spectral and temporal properties because the spectral properties of $\ket{n_k}_{\phi_k}$ differ from those of $\ket{n_k}_{\varphi_k}$.

The states in Eq. (\ref{eq:phi_k}) can be written more succinctly  as (see Appendix \ref{gibbsform})
\begin{align}\label{gibbs}
\rho_{\phi_k} ={}& \frac{1}{Z_k}e^{-\alpha_k \hat A_k^\dagger \hat A_k }\,;\\
Z_k ={}& \text{Tr}\left( e^{-\alpha_k \hat A_k^\dagger \hat A_k }\right)= \frac{1}{1-e^{-\alpha_k }}\,,
\end{align}
where $Z_k$ is the partition function of mode $k$ and 
\begin{align}
e^{-\alpha_k} = \tanh^2(r_k) = \frac{\bar{n}_k}{1+\bar{n}_k}\,.
\end{align}
Using this notation we can also write the full state in mode $a$  as
\begin{align}
\rho_a &=\frac{1}{Z}e^{-\sum_k\alpha_k \hat A_k^\dagger \hat A_k }\,;\\
Z&=  \text{Tr}\left( e^{-\sum_k\alpha_k \hat A_k^\dagger \hat A_k }\right)\,.
\end{align}

\section{Properties of broadband pseudothermal states}
To study the coherence properties of the  broadband states, we compute various  temporal  and spectral correlation functions.  The expressions for mode $a$ can be mapped to those for mode $b$ by making the substitution: $\phi_k\rightarrow\varphi_k$.

\subsection{Spectral correlation function}
To study the spectral coherence properties of the light, we introduce a spectral correlation function $S(\omega,\omega')$. Using the procedure outlined in Appendix \ref{nicoapp}, one finds that
\begin{align}
S_a(\omega,\omega')={}&\braket{\hat a^\dagger(\omega) \hat a(\omega')}_\psi \\
={}&\sum_ k \sinh^2(r_k) \phi_k^*(\omega) \phi_k(\omega')\,,
\end{align}
and also that $\braket{\hat a(\omega) \hat a(\omega')}_\psi=\braket{\hat a^\dagger(\omega) \hat a^\dagger(\omega')}_\psi =0$. 

The same-frequency correlation function is 
\begin{align}
S_a(\omega,\omega)=\sum_ k |\phi_k(\omega)|^2 \bar{n}_k\,,
\end{align}
which can be interpreted as the spectral density. 

\subsection{First-order temporal correlation function}

The  first-order temporal correlation function for mode $a$ is   \cite{Glauber1963c}:
\begin{align}
G_a^{(1)}(t_1,t_2)=\left\langle\hat{a}^{\dagger}(t_1)\hat{a}(t_2)\right\rangle_{\psi}\,.
\end{align}
In the limit of the state $\ket{\psi}$ having sufficiently narrow frequency support this quantity is proportional to $\langle\hat{E}^{(-)}(t_1)\hat{E}^{(+)}(t_2)\rangle_{\psi}\,$ where $\hat{E}_a^{(\pm)}(t_1)$ are the usual positive or negative frequency components of the electric field operator in mode $a$. The operators $\hat a(t)$ are nothing but the Fourier transform of the operators $\hat a(\omega)$ 
\begin{align}
\hat{a}(t)=\frac{1}{\sqrt{2\pi}}\int d\omega \hat{a}(\omega)e^{i\omega t}\,.
\end{align}
We thus have
\begin{align}
G_a^{(1)}(t_1,t_2)={}&\left\langle\hat{a}\dg(t_1)\hat{a}(t_2)\right\rangle_{\psi}\\
\label{eq:G1a}
={}&\int \frac{d\omega d\omega'}{2 \pi} S(\omega,\omega')e^{-i \omega t_1 +i\omega' t_2}\\\label{eq:cc}
={}&\sum_k \tilde \phi^*_k(t_1)\tilde \phi_k(t_2)\bar{n}_k\,,
\end{align}
where $\bar{n}_k$ is defined in Eq. (\ref{eq:nk}), and where
\begin{align}\label{eq:tildephi}
\tilde \phi_k(t)=\frac{1}{\sqrt{2\pi}}\int d\omega \phi_k(\omega)e^{i\omega t}\,.
\end{align}
This expression is derived in Appendix \ref{sec:der}. To see the temporal distribution, we compute
\begin{align}\label{eq:nnm}
G_a^{(1)}(t,t)=\sum_k |\tilde \phi_k(t)|^2\bar{n}_k\,,
\end{align}
which can be interpreted as the probability per unit time that a photon is absorbed by an ideal detector at time $t$  \cite{Glauber1963c}.

We can also compute a normalized first-order correlation function
\begin{align}
g^{(1)}(t_1,t_2)=\frac{G^{(1)}(t_1,t_2)}{\sqrt{G^{(1)}(t_1,t_1)G^{(1)}(t_2,t_2)}}\,.
\end{align}
In the special case where the decomposition has only one Schmidt mode, the normalized first order correlation function is $|g^{(1)}(t_1,t_2)|=1$. In other words, this is a state with finite bandwidth and thermal photon-number statistics, and yet it has perfect first-order coherence \cite{Glauber1963c}.  

It is known from classical coherence theory that correlation functions for partially coherent light can be decomposed as sums of coherent mode functions \cite{Mandel1995}.  The connection between such spatial coherent modes and spatial Schmidt modes in SPDC was suggested and explored experimentally in \cite{Bobrov2013}. Here, we show this explicitly,  in terms of spectral coherent modes. Eq. (\ref{eq:cc}) shows that for a single beam of a twin-beam state with arbitrary spectral entanglement, these mode functions are indeed the (spectral) Schmidt modes.

\subsection{Second-order temporal correlation function}
To see intensity-intensity correlations, we look at the  second-order correlation function \cite{Glauber1963c}:
\begin{align}
G^{(2)}(t_1,t_2)={}&\left\langle\hat{a}\dg(t_1)\hat{a}\dg(t_2)\hat{a}(t_1)\hat{a}(t_2)\right\rangle_{\psi}\\\label{eq:g2a}
\begin{split}
={}&G_a^{(1)}(t_1,t_1)G_a^{(1)}(t_2,t_2)\\
&+G_a^{(1)}(t_1,t_2)G_a^{(1)}(t_2,t_1)\,,
\end{split}
\end{align}
where we again replaced the electric field operators with photon number creation and destruction operators, and where $G_a^{(1)}(t_1,t_2)$ is defined in Eq. (\ref{eq:G1a}). This expression is derived in Appendix \ref{sec:der}.
Eq. (\ref{eq:g2a}) can be interpreted as the probability per unit $(\mathrm{time})^2$ that one photon is recorded at time $t_1$ and another at time $t_2$ \cite{Glauber1963c}.

\section{Continuous Wave limit}

In the case of a cw laser driving an SPDC process at frequency $\bar{\omega}_p$ (or a SFWM process at $\bar{\omega}_p/2$), energy is conserved according to $\omega_a+\omega_b=\bar{\omega}_p$. The two-mode squeezed state has the form
\begin{align}
 \ket{\psi} = {}&\mathcal{\hat U}_{\text{SQ}} \ket{\text{vac}}\,;\\
 \mathcal{\hat U}_{\text{SQ}} ={}  & e^{ \left(\int \mathrm d \omega\, r(\omega) \hat{a}^\dagger(\omega) \hat{b}^\dagger(\bar \omega_p -\omega) - \hc \right) }\,.
\end{align}
We can also construct  Heisenberg picture transformation (similar to the pulsed-pump case):
\begin{align}
\hat a(\omega) &\to{} \mathcal{\hat U}_{\text{SQ}}^\dagger \hat a(\omega)  \mathcal{\hat U}_{\text{SQ}}\\
={}& \cosh(r(\omega)) \hat a(\omega) +\sinh(r(\omega)) b^\dagger(\omega_p{-}\omega)\\
\hat b(\omega) &\to {}\mathcal{\hat U}_{\text{SQ}}^\dagger \hat b(\omega)  \mathcal{\hat U}_{\text{SQ}} \\
={}& \cosh(r(\omega)) \hat b(\omega)+\sinh(r(\omega)) a^\dagger(\omega_p{-}\omega)\,,
\end{align}
and also write states for mode $a$ (or $b$)
\begin{align}
\rho_a &= \frac{1}{Z}e^{- \int \mathrm d \omega \alpha(\omega) \hat a^\dagger (\omega) \hat a(\omega) }\,,\\
Z&=\text{Tr}\left(e^{- \int \mathrm d \omega \alpha(\omega) \hat a^\dagger (\omega) \hat a(\omega) }\right)\,,
\end{align}
where $\alpha(\omega) = \ln ( 1/ \tanh^{2}(r (\omega)) )$.
By generalizing the result in  Appendix \ref{nicoapp}, from a discrete to a continuum index, the spectral correlation function becomes 
\begin{align}\label{eq:Scw}
S_{\textsc{cw}}(\omega,\omega') &=\bar{n}_{\textsc{cw}}(\omega) \delta(\omega-\omega')\,,\\
\bar{n}_{\textsc{cw}}(\omega)& = \sinh^2(r(\omega))\,.
\end{align}
Eq. (\ref{eq:Scw}) tells us that there are no spectral correlations between two different frequency modes.  

The first-order correlation function for  cw broadband thermal light is the two-dimensional Fourier transform of  $S_{\textsc{cw}}(\omega,\omega')$. After integrating out the Dirac delta function, this becomes
 \begin{align}\label{eq:g1cw}
G_{\textsc{cw}}^{(1)}(t_1,t_2)={}&\frac{1}{2\pi}\int \mathrm d  \omega e^{i\omega(t_1-t_2)}\bar{n}_{\textsc{cw}}(\omega)\,,
\end{align}
as expected  from the Wiener-Khinchin theorem. From this, we can compute:
\begin{align}\label{eq:nnm2}
G_{\textsc{cw}}^{(1)}(t,t)=\frac{1}{2\pi}\int \mathrm d  \omega  \bar{n}_{\textsc{cw}}(\omega)\,.
\end{align}
That  Eq. (\ref{eq:nnm2}) is constant also shows that the light is fully spectrally incoherent, since time-varying intensities arise from well-defined phases between frequencies. Similarly, the second-order correlation function for cw-pumped BPT states is 
\begin{align}\label{eq:g2cw}
\begin{split}
G^{(2)}_{\textsc{cw}}(t_1,t_2)={}&G_{\textsc{cw}}^{(1)}(t_1,t_1)G_{\textsc{cw}}^{(1)}(t_2,t_2)\\
&+G_{\textsc{cw}}^{(1)}(t_1,t_2)G_{\textsc{cw}}^{(1)}(t_2,t_1)\,,
\end{split}
\end{align}
where $G_{\textsc{cw}}^{(1)}(t_1,t_2)$ is defined in Eq. (\ref{eq:g1cw}).

If  one sets $\alpha(\omega)=\hbar\omega/k_B T$, then $\rho_a $ can be thought of as a multi-mode thermal state in the traditional sense. Eq. (\ref{eq:g2cw}) then becomes identical to the expression for $G^{(2)}(t_1,t_2)$ for chaotic light, derived by, e.g.,  Loudon \cite{Loudon1983}. The shape of $\alpha(\omega)$ depends on $r(\omega)$ which in turn depends on the nonlinearity profile of the material. It would therefore be difficult to make light that exactly matches the Planck spectrum. But one could match the Planck spectrum over a finite bandwidth, or make more general states where each frequency corresponds to a thermal state at a different temperature $T$.

\section{Examples}
To illustrate the pump laser's impact on the  coherence of BPT states, we examine the concrete case of SPDC. We  look at three examples: short pulse; long pulse; and cw. 

For simplicity, we consider pump fields with Gaussian spectral distributions (e.g., shaped using optical pulse-shaping \cite{Weiner2011}), and crystals with Gaussian nonlinearity profiles (e.g., engineered using  nonlinearity shaping methods \cite{Branczyk2011,Dixon2013,Dosseva2016,Tambasco2016,Graffitti2017}). We consider the system to be in the low-gain regime and satisfy symmetric group-velocity-matching \cite{ansari2018tailoring,kuzucu2008joint}. 

\begin{figure*}[h!]
\begin{center}
\includegraphics[width=\textwidth]{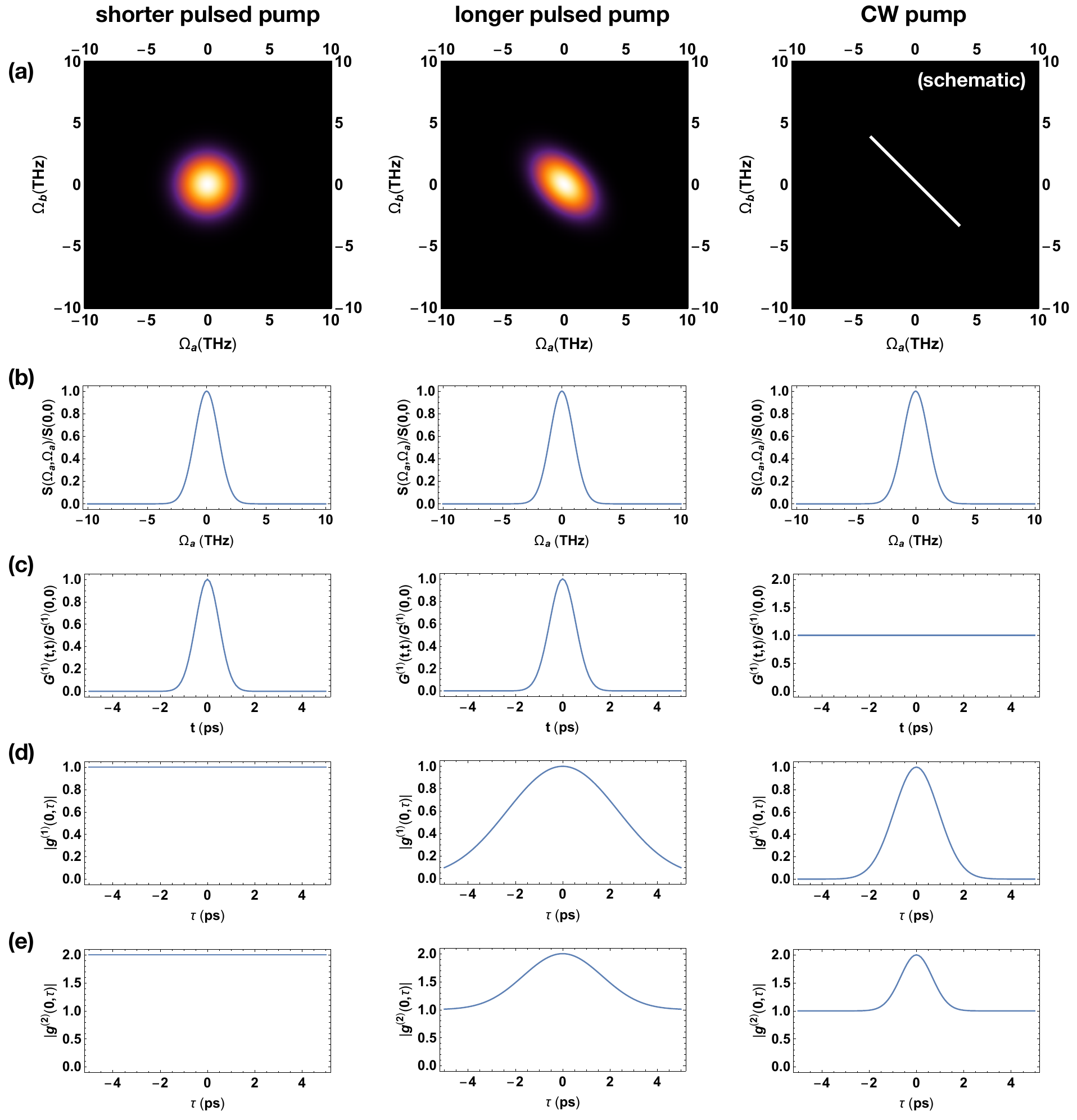}
\end{center}
\caption{ Comparison between sources pumped by a shorter-pulsed, longer-pulsed and CW laser: (a) the joint spectral amplitude $J(\omega_1,\omega_2)$; (b) equal-frequency spectral correlation function (normalized), (c) equal-time first-order temporal correlation function (normalized); (d) time-separated normalized first-order temporal correlation function; and (e)  time-separated normalized second-order temporal correlation function $g^{(2)}(t_1,t_2)=1+g^{(1)}(t_1,t_2)g^{(1)}(t_2,t_1)$. Notice that only the cw-pumped case satisfies  the Wiener-Khinchin theorem. Also, notice that the time-integrated $g^{(2)}$ decreases with increasing number of Schmidt modes, as was shown in \cite{Christ2011}. }
\label{fig:jsa}
\end{figure*}

We take
\begin{align}
J(\Omega_a,\Omega_b)=Ae^{-\frac{(\Omega_a+\Omega_b)^2}{2\sigma_p^2}}e^{-\frac{(\Omega_a-\Omega_b)^2}{2\sigma_c^2}}
\end{align}
for the pulsed case, where $\Omega_j=\omega_j-\bar{\omega}_j$ (with $\bar{\omega}_j$ the mean frequency of the downconverted beam), $\sigma_p$ is the spectral width of the pump amplitude function, and $\sigma_c$ is the width of the phase-matching function (given as the Fourier transform of the  longitudinal shape of the material nonlinearity). We also take
\begin{align}
r(\Omega)=Ae^{-\frac{(2\Omega)^2}{2\sigma_c^2}}\,,
\end{align}
for the cw case, where $\Omega=\omega-\bar{\omega}_a$. Figure \ref{fig:jsa} compares  three examples.  For easy comparison, we chose $\sigma_p$ and $\sigma_c$ to make BPT states with similar spectral distributions  (the choice of  $A$ does not affect the normalized functions). These are shown in Table \ref{tab:params}.

\begin{table}[h!]
\begin{tabular}{|c||c|c|}\hline  pump & ~~$\sigma_p$~~ &~~ $\sigma_c$ ~~\\\hline shorter &$2\times 10^{12}$ s$^{-1}$  & $2\times 10^{12}$  s$^{-1}$   \\ longer &   $1.5\times 10^{12}$ s$^{-1}$  & $2.4\times 10^{12}$ s$^{-1}$  \\ CW  &  n/a &  $3\times 10^{12}$  s$^{-1}$ \\ \hline \end{tabular}
\caption{Parameters used to generate Figure \ref{fig:jsa}. }
\label{tab:params}
\end{table}

Figure \ref{fig:jsa} (a) shows the JSA of the twin-beam state. The JSA for the shorter-pulse case is separable (note that this only happens for a pair of \emph{Gaussians} of the same width \cite{quesada2018gaussian}). The longer-pulsed laser  leads to a slightly correlated JSA, and the cw laser  leads to a strongly correlated JSA. 

Figure \ref{fig:jsa} (b) shows the spectral distributions in mode $a$, chosen to be the same for all cases. 

Figure \ref{fig:jsa} (c) shows the temporal distributions in mode $a$. These vary between the three  examples. Because the short pump  yields a separable JSA, the BPT state is prepared in a single, broad, yet coherent, spectral mode. This means that the frequencies  have fixed relative phases, which leads to a spectrally-coherent pulse of pseudothermal light. As the pump pulse-length increases slightly,  the pseudothermal pulse gets slightly longer. In turn, the relative phases of the frequencies become slightly less fixed. In the limit of a cw pump, the BPT state is also continuous, but broadband, and the frequencies have \emph{no} fixed relative phase relationship. 

\clearpage

Figure \ref{fig:jsa} (d) shows the normalized time-resolved first-order temporal correlation function  in mode $a$. The light in all three examples has drastically different coherence times despite having very similar spectra. For a separable JSA, $|g^{(1)}(t_1,t_2)|=1$, but as the spectral correlations increase, the coherence time decreases.  We note that since a pulse is not stationary, it does not have to satisfy the Wiener-Khinchin theorem. 

Figure \ref{fig:jsa} (e) shows the normalized time-resolved second-order temporal correlation function in mode $a$. For a separable JSA, $|g^{(2)}(t_1,t_2)|=2$, but as the spectral correlations increase, the second-order coherence time decreases. 

In summary, these examples show that by varying the spectrum of the pump field as well as the shape of the nonlinearity function of the material, it is possible to generate various states of light with the same spectrum,  but drastically different spectral, and first- and second-order temporal coherence properties.

\section{Discussion}

Broadband states generated via nonlinear optical processes, such as SPDC or SFWM, have interesting coherence properties. We calculated their spectral and temporal  correlation functions in terms of the Schmidt modes of the joint spectral amplitude, and showed that these states can be thought of as pseudothermal states. We also showed that these states have tuneable spectral coherence---which can be tuned from perfectly spectrally coherent to completely incoherent. Regardless of the level of spectral coherence, these states can be decomposed into tensor products of states with geometric photon-number statistics.  

High-gain SPDC can be extremely bright for a parametric process---up to hundreds of mW mean power \cite{Perez2015,Spasibko2016,Spasibko2017}. This is close to the $\sim$1 W achievable via non-parametric processes \cite{Ulness2003}, and is  bright enough to study interesting non-classical phenomena such as the interplay between spectral and photon-number coherence. Furthermore, a source based on SPDC is extremely customizable. Beyond tuning the spectral coherence, the spectral shape can be customized using optical pulse-shaping   \cite{Weiner2011} or  nonlinearity shaping methods \cite{Branczyk2011,Dixon2013,Dosseva2016,Tambasco2016,Graffitti2017}. 

Broadband spectrally incoherent light generated by nonlinear optics may  also have application in studying the dynamics of photoinduced processes, such as the time scales and mechanisms underlying the initial step of photosynthesis in light-harvesting complexes. Some researchers have  questioned whether dynamics initiated by sunlight excitation might be different from those detected in femtosecond laser experiments performed on light-harvesting complexes \cite{Jiang1991,Mancal2010,Hoki2011,Brumer2012,Kassal2013}. A broadband source with tuneable spectral coherence might  help answer this question. 

On the theory side, this work is relevant to the study of decompositions of thermal light into broadband coherent modes. Thermal light cannot be represented as a statistical mixture of single pulses \cite{Chenu2015}, but  one can construct mixtures of single pulses that yield the same first-order temporal correlation function as thermal light \cite{Chenu2015b}. In a one-dimensional waveguide, thermal light was shown to decompose into a statistical mixture of sets of coherent pulses \cite{Branczyk2017}. Here, we show that partially-spectrally-coherent  light can be decomposed into a tensor product of states prepared in spectral Schmidt modes, each with geometric photon-number statistics. Our work also connects with decompositions, into Schmidt-like modes, of correlation functions for partially spatially coherent thermal light \cite{Bobrov2013}.

Our formalism also provides a convenient tool for computing any observables of broadband pseudothermal states with partial spectral coherence---even for states created with completely different experimental methods, such as \cite{Schulz2005,Jechow2013,Hartmann2015,Mika2018,Zhou2018}. This is due to a well-known result in quantum information theory, the Stinespring dilation theorem \cite{Stinespring1955},  that states that any mixed state can be represented as a pure state in a higher-dimensional Hilbert space. Any broadband pseudothermal state can therefore be written as the reduced state of a hypothetical broadband twin-beam state. 

The relationship between spectral, temporal, and photon-number coherence is of fundamental and practical interest. We hope that our analysis of broadband pseudothermal states generated via nonlinear optics provides a useful way of exploring it theoretically and experimentally.

\section*{Acknowledgements}
The authors thank John Sipe for helpful discussions, and John Donohue for helpful comments on the manuscript. 

\section*{Funding Information}
 Research at Perimeter Institute is supported by the Government of Canada through Industry Canada and by the Province of Ontario through the Ministry of Research and Innovation. We acknowledge the support of the Natural Sciences and Engineering Research Council of Canada. 


\begin{thebibliography}{10}

\bibitem{Zavatta2007}
A. Zavatta, V. Parigi, and M. Bellini, ``Experimental nonclassicality of
  single-photon-added thermal light states,'' Phys. Rev. A {\bf 75}, 052106
  (2007).

\bibitem{Tahira2009}
R. Tahira {\it et~al.}, ``Entanglement of Gaussian states using a beam
  splitter,'' Phys. Rev. A {\bf 79}, 023816 (2009).

\bibitem{Kim2002a}
M.~S. Kim, ``Entanglers: Beam splitters and Thermal fields,'' Fortschritte der
  Physik {\bf 50}, 652 (2002).

\bibitem{Weedbrook2012}
C. Weedbrook, S. Pirandola, and T.~C. Ralph, ``Continuous-variable quantum key
  distribution using thermal states,'' Phys. Rev. A {\bf 86}, 022318 (2012).

\bibitem{Harder2014}
G. Harder {\it et~al.}, ``Tomography by Noise,'' Phys. Rev. Lett. {\bf 113},
  070403 (2014).

\bibitem{Gatti2004a}
A. Gatti {\it et~al.}, ``Correlated imaging, quantum and classical,'' Phys.
  Rev. A {\bf 70}, 013802 (2004).

\bibitem{Gatti2004}
A. Gatti {\it et~al.}, ``Ghost Imaging with Thermal Light: Comparing
  Entanglement and Classical Correlation,'' Phys. Rev. Lett. {\bf 93}, 093602
  (2004).

\bibitem{Valencia2005}
A. Valencia {\it et~al.}, ``Two-Photon Imaging with Thermal Light,'' Phys. Rev.
  Lett. {\bf 94}, 063601 (2005).

\bibitem{Zhang2005b}
D. Zhang {\it et~al.}, ``Correlated two-photon imaging with true thermal
  light,'' Opt. Lett. {\bf 30}, 2354 (2005).

\bibitem{Agafonov2009}
I.~N. Agafonov {\it et~al.}, ``High-visibility intensity interference and ghost
  imaging with pseudo-thermal light,'' Journal of Modern Optics {\bf 56}, 422
  (2009).

\bibitem{Chen2010}
X.-H. Chen {\it et~al.}, ``High-visibility, high-order lensless ghost imaging
  with thermal light,'' Opt. Lett. {\bf 35}, 1166 (2010).

\bibitem{Moreau2017}
P.-A. Moreau {\it et~al.}, ``Ghost Imaging Using Optical Correlations,''
  {L}aser {P}hotonics {R}ev \ p.\ 1700143 (2017).

\bibitem{Cao2010}
D.-Z. Cao, G.-J. Ge, and K. Wang, ``Two-photon subwavelength lithography with
  thermal light,'' Applied Physics Letters {\bf 97}, 051105 (2010).

\bibitem{Lu2008}
G. Lu {\it et~al.}, ``Improving Visibility of Diffraction Pattern with
  Pseudo-Thermal Light,'' Chinese Physics Letters {\bf 25}, 1277 (2008).

\bibitem{Sprigg2016}
J. {Sprigg}, T. {Peng}, and Y. {Shih}, ``{Super-resolution imaging using the
  spatial-frequency filtered intensity fluctuation correlation},'' Sci. Rep.
  (2016).

\bibitem{Lajunen2009}
H. Lajunen {\it et~al.}, ``Resolution-enhanced optical coherence tomography
  based on classical intensity interferometry,'' J. Opt. Soc. Am. A {\bf 26},
  1049 (2009).

\bibitem{Lopez-Mariscal2010}
C. L{\'o}pez-Mariscal and J.~C. Guti{\'e}rrez-Vega, ``Observation of optical
  guiding using thermal light,'' Journal of Optics {\bf 12}, 075702 (2010).

\bibitem{Ulness2003}
D.~J. Ulness, ``On the Role of Classical Field Time Correlations in Noisy Light
  Spectroscopy: Color Locking and a Spectral Filter Analogy,'' J. {C}hem.
  {P}hys. {\bf 107}, 8111 (2003).

\bibitem{Turner2013}
D.~B. Turner {\it et~al.}, ``Coherent multidimensional optical spectra measured
  using incoherent light,'' Nat Commun 4 (2013).

\bibitem{Li2005}
G. Li {\it et~al.}, ``Correction of photon statistics of quantum states in
  single-photon detection,'' in {\it Quantum Optics and Applications in
  Computing and Communications II}, {\bf 5631}, 134  (2005).

\bibitem{Iskhakov2011}
T. Iskhakov {\it et~al.}, ``Intensity correlations of thermal light,'' The
  European Physical Journal Special Topics {\bf 199}, 127 (2011).

\bibitem{Zhu2012a}
J. Zhu {\it et~al.}, ``Thermal-light-based ranging using second-order
  coherence,'' Applied optics {\bf 51}, 4885 (2012).

\bibitem{Jechow2013}
A. Jechow {\it et~al.}, ``Enhanced two-photon excited fluorescence from imaging
  agents using true thermal light,'' Nature Photonics {\bf 7}, 973 EP  (2013).

\bibitem{Hartmann2015}
S. Hartmann {\it et~al.}, ``Tailored quantum statistics from broadband states
  of light,'' New {J}. {P}hys. {\bf 17}, 043039 (2015).

\bibitem{Mika2018}
J. {Mika} {\it et~al.}, ``{Broadband thermal light with Bose-Einstein photon
  statistics from warm atomic vapor},'' arXiv:1801.10555 [quant-ph]  (2018).

\bibitem{Schulz2005}
T.~F. Schulz {\it et~al.}, ``Complete cancellation of noise by means of color-
  locking in nearly degenerate, four-wave mixing of noisy light,'' {J}. {O}pt.
  {S}oc. {A}m. {B} {\bf 22}, 1052 (2005).

\bibitem{Christ2011}
A. Christ {\it et~al.}, ``Probing multimode squeezing with correlation
  functions,'' New Journal of Physics {\bf 13}, 033027 (2011).

\bibitem{Spasibko2017}
K.~Y. Spasibko {\it et~al.}, ``Multiphoton Effects Enhanced due to Ultrafast
  Photon-Number Fluctuations,'' Phys. Rev. Lett. {\bf 119}, 223603 (2017).

\bibitem{Vernon2017}
Z. Vernon {\it et~al.}, ``Truly unentangled photon pairs without spectral
  filtering,'' Optics letters {\bf 42}, 3638 (2017).

\bibitem{Blauensteiner2009}
B. Blauensteiner {\it et~al.}, ``Photon bunching in parametric down-conversion
  with continuous-wave excitation,'' Phys. Rev. A {\bf 79}, 063846 (2009).

\bibitem{eckstein2011highly}
A. Eckstein {\it et~al.}, ``Highly efficient single-pass source of pulsed
  single-mode twin beams of light,'' Physical Review Letters {\bf 106}, 013603
  (2011).

\bibitem{Spasibko2012}
K.~Y. Spasibko, T.~S. Iskhakov, and M.~V. Chekhova, ``Spectral properties of
  high-gain parametric down-conversion,'' Opt. Express {\bf 20}, 7507 (2012).

\bibitem{Wakui2014}
K. Wakui {\it et~al.}, ``Ultrabroadband direct detection of nonclassical photon
  statistics at telecom wavelength,'' Scientific Reports {\bf 4}, 4535 EP
  (2014).

\bibitem{Huang1989}
J. Huang and P. Kumar, ``Photon-counting statistics of multimode squeezed
  light,'' Phys. Rev. A {\bf 40}, 1670 (1989).

\bibitem{ansari2018tomography}
V. Ansari {\it et~al.}, ``Tomography and purification of the temporal-mode
  structure of quantum light,'' Physical review letters {\bf 120}, 213601
  (2018).

\bibitem{Grice1997}
W.~P. Grice and I.~A. Walmsley, ``Spectral information and distinguishability
  in type-II down-conversion with a broadband pump,'' Phys. Rev. A {\bf 56},
  1627 (1997).

\bibitem{Quesada2014}
N. Quesada and J.~E. Sipe, ``Effects of time ordering in quantum nonlinear
  optics,'' Phys. Rev. A {\bf 90}, 063840 (2014).

\bibitem{Quesada2015}
N. Quesada and J.~E. Sipe, ``Time-Ordering Effects in the Generation of
  Entangled Photons Using Nonlinear Optical Processes,'' Phys. Rev. Lett. {\bf
  114}, 093903 (2015).

\bibitem{quesada2018b}
{N. Quesada, G. Triginer, M. D. Vidrighin, and J.~E. {Sipe}}, In preparation.

\bibitem{mehler1866ueber}
F.~G. Mehler, ``Ueber die Entwicklung einer Function von beliebig vielen
  Variablen nach Laplaceschen Functionen h{\"o}herer Ordnung.,'' Journal
  f{\"u}r die reine und angewandte Mathematik {\bf 66}, 161 (1866).

\bibitem{Graffitti2018b}
F. Graffitti {\it et~al.}, ``Design considerations for high-purity heralded
  single-photon sources,'' Physical Review A {\bf 98}, 053811 (2018).

\bibitem{Glauber1963c}
R.~J. Glauber, ``The Quantum Theory of Optical Coherence,'' Phys. Rev. {\bf
  130}, 2529 (1963).

\bibitem{Mandel1995}
L. Mandel and E. Wolf, {\it Optical coherence and quantum optics} (Cambridge
  university press, 1995).

\bibitem{Bobrov2013}
I.~B. Bobrov {\it et~al.}, ``Schmidt-like coherent mode decomposition and
  spatial intensity correlations of thermal light,'' New Journal of Physics
  {\bf 15}, 073016 (2013).

\bibitem{Loudon1983}
R. Loudon, {\it The {Q}uantum {T}heory of {L}ight} (Clarendon Press, Oxford,
  1983).

\bibitem{Weiner2011}
A.~M. Weiner, ``Ultrafast optical pulse shaping: A tutorial review,'' Optics
  Communications {\bf 284}, 3669 (2011).

\bibitem{Branczyk2011}
A.~M. Bra\'{n}czyk {\it et~al.}, ``Engineered optical nonlinearity for quantum
  light sources,'' Opt. Express {\bf 19}, 55 (2011).

\bibitem{Dixon2013}
P.~B. Dixon, J.~H. Shapiro, and F.~N.~C. Wong, ``Spectral engineering by
  Gaussian phase-matching for quantum photonics,'' Opt. Express {\bf 21}, 5879
  (2013).

\bibitem{Dosseva2016}
A. Dosseva, L. Cincio, and A.~M. Bra\ifmmode~\acute{n}\else \'{n}\fi{}czyk,
  ``Shaping the joint spectrum of down-converted photons through optimized
  custom poling,'' Phys. Rev. A {\bf 93}, 013801 (2016).

\bibitem{Tambasco2016}
J.-L. Tambasco {\it et~al.}, ``Domain engineering algorithm for practical and
  effective photon sources,'' Opt. Express {\bf 24}, 19616 (2016).

\bibitem{Graffitti2017}
F. Graffitti {\it et~al.}, ``Pure down-conversion photons through
  sub-coherence-length domain engineering,'' Quantum Science and Technology
  {\bf 2}, 035001 (2017).

\bibitem{ansari2018tailoring}
V. Ansari {\it et~al.}, ``Tailoring nonlinear processes for quantum optics with
  pulsed temporal-mode encodings,'' Optica {\bf 5}, 534 (2018).

\bibitem{kuzucu2008joint}
O. Kuzucu {\it et~al.}, ``Joint temporal density measurements for two-photon
  state characterization,'' Physical review letters {\bf 101}, 153602 (2008).

\bibitem{quesada2018gaussian}
N. Quesada and A.~M. Bra\ifmmode~\acute{n}\else \'{n}\fi{}czyk, ``Gaussian
  functions are optimal for waveguided nonlinear-quantum-optical processes,''
  Phys. Rev. A {\bf 98}, 043813 (2018).

\bibitem{Perez2015}
A.~M. P{\'e}rez {\it et~al.}, ``Giant narrowband twin-beam generation along the
  pump-energy propagation direction,'' Nature communications {\bf 6}, 7707
  (2015).

\bibitem{Spasibko2016}
K.~Y. Spasibko {\it et~al.}, ``Ring-shaped spectra of parametric downconversion
  and entangled photons that never meet,'' Optics letters {\bf 41}, 2827
  (2016).

\bibitem{Jiang1991}
X.-P. Jiang and P. Brumer, ``Creation and dynamics of molecular states prepared
  with coherent vs partially coherent pulsed light,'' The Journal of Chemical
  Physics {\bf 94}, 5833 (1991).

\bibitem{Mancal2010}
T. Man{\v c}al and L. Valkunas, ``Exciton dynamics in photosynthetic complexes:
  excitation by coherent and incoherent light,'' New Journal of Physics {\bf
  12}, 065044 (2010).

\bibitem{Hoki2011}
K. Hoki and P. Brumer, ``Excitation of Biomolecules by Coherent vs. Incoherent
  Light: Model Rhodopsin Photoisomerization,'' Procedia Chemistry {\bf 3}, 122
  (2011).

\bibitem{Brumer2012}
P. Brumer and M. Shapiro, ``Molecular response in one-photon absorption via
  natural thermal light vs. pulsed laser excitation,'' Proceedings of the
  National Academy of Sciences {\bf 109}, 19575 (2012).

\bibitem{Kassal2013}
I. Kassal, J. Yuen-Zhou, and S. Rahimi-Keshari, ``Does Coherence Enhance
  Transport in Photosynthesis?,'' The Journal of Physical Chemistry Letters
  {\bf 4}, 362 (2013).

\bibitem{Chenu2015}
A. Chenu {\it et~al.}, ``Thermal Light Cannot Be Represented as a Statistical
  Mixture of Single Pulses,'' Phys. Rev. Lett. {\bf 114}, 213601 (2015).

\bibitem{Chenu2015b}
A. Chenu, A.~M. Bra\ifmmode~\acute{n}\else \'{n}\fi{}czyk, and J.~E. Sipe,
  ``First-order decomposition of thermal light in terms of a statistical
  mixture of single pulses,'' Phys. Rev. A {\bf 91}, 063813 (2015).

\bibitem{Branczyk2017}
A.~M. Bra\'{n}czyk, A. Chenu, and J.~E. Sipe, ``Thermal light as a mixture of
  sets of pulses: the quasi-1D example,'' J. Opt. Soc. Am. B {\bf 34}, 1536
  (2017).

\bibitem{Zhou2018}
Y. {Zhou} {\it et~al.}, ``{Superbunching pseudothermal light with intensity
  modulated laser light and rotating groundglass},'' ArXiv e-prints  (2018).

\bibitem{Stinespring1955}
W.~F. Stinespring, ``Positive functions on C*-algebras,'' Proceedings of the
  American Mathematical Society {\bf 6}, 211 (1955).

\end{thebibliography}

\onecolumngrid
\appendix

\section{Writing the single Schmidt mode thermal state in Gibbs form}\label{gibbsform}
Consider the number operator of the $k^{\text{th}}$ Schmidt mode 
\begin{align}
\hat n_k = \hat A_k^\dagger \hat A_k.
\end{align}
We can write its eigendecomposition and the resolution of the identity as follows
\begin{align}
\hat n_k \ket{n_k} = n_k \ket{n_k}, \quad \mathbb{I}_k = \sum_{n_k=0}^\infty \ket{n_k} \bra{n_k}.
\end{align}
Using these expressions we  calculate
\begin{align}
\exp(-\alpha_k \hat n_k) = \exp(-\alpha_k \hat A_k^\dagger \hat A_k) = \sum_{n_k=0}^\infty e^{-\alpha _k n_k} 	\ket{n_k} \bra{n_k}, \\
\text{Tr}\left(\exp(-\alpha_k \hat n_k)  \right) = \sum_{n_k=0}^\infty e^{-\alpha _k n_k}  = \frac{1}{1-e^{-\alpha_k}},
\end{align}
where in the last expression we use the geometric series sum. Consider now their ratio
\begin{align}
\frac{\exp(-\alpha_k \hat A_k^\dagger \hat A_k)	}{\text{Tr}\left(\exp(-\alpha_k \hat A_k^\dagger \hat A_k) \right)} = \sum_{n_k=0}^\infty \left(1-e^{-\alpha_k} \right) \left(e^{-\alpha_k} \right)^{n_k} \ket{n_k} \bra{n_k},
\end{align}
We can compare this with Eq. (\ref{gibbs0}) and Eq. (\ref{gibbs1}) and identify
\begin{align}
e^{-\alpha_k} = \frac{\bar{n}_k}{\bar{n}_k+1}	,
\end{align}
easily verifying that $1-e^{-\alpha_k} = 1/(\bar{n}_k+1)$, completing the derivation of Eq. (\ref{gibbs}).

\section{Derivation of $G^{(1)}(\omega,\omega')$ and $G^{(2)}(\omega,\omega')$}\label{nicoapp}

In this section, we  derive the frequency second- and fourth-order moments of the $a$ fields. For the second-order moment, we use the relations in Eq. (\ref{eq:invert}) to write
\begin{align}\label{exp}
\braket{\hat{a}^\dagger(\omega) \hat{a}(\omega')}_\psi = \sum_{k l } \phi_k^*(\omega) \phi_l(\omega) \braket{A_k^\dagger A_l}_\psi\,.
\end{align}
We then use the linear Bogoliubov transformations in Eqs. (\ref{HeisenbergEq}) to write
\begin{align}
\braket{A_k^\dagger A_l}_\psi &= \braket{\text{vac}|\mathcal{\hat U}_{\text{SQ}}^\dagger \hat{A}_k  \mathcal{\hat U}_{\text{SQ}} \mathcal{\hat U}_{\text{SQ}}^\dagger \hat{A}_l  \mathcal{\hat U}_{\text{SQ}} |\text{vac}} \\
&= \braket{\text{vac}|\left(\cosh(r_k) \hat{A}_k+ \sinh(r_k) \hat{B}_k^\dagger \right) \left( \cosh(r_l) \hat{A}_l+ \sinh(r_l) \hat{B}_l^\dagger \right) |\text{vac}}\\
&=\sinh(r_k) \sinh(r_l) \braket{\text{vac}|\hat B_k \hat B_l^\dagger |\text{vac}}\\\label{eq:thing}
&=\sinh^2(r_k) \delta_{k,l}\,.
\end{align}
Plugging this  result into Eq. (\ref{exp}), we obtain
\begin{align}\label{g1nico}
\braket{\hat{a}^\dagger(\omega) \hat{a}(\omega')}_\psi = \sum_{ k } \sinh^2(r_k)\phi_k^*(\omega) \phi_k(\omega) \,.
\end{align}
Now let us consider the fourth-order moment
\begin{align}\label{G2nico}
\braket{\hat a^\dagger (\omega) \hat a^\dagger (\omega') \hat a(\omega) \hat a(\omega')}_\psi = \sum_{k,l,m,n} \phi_k(\omega) \phi_l(\omega') \phi^*_m(\omega) \phi^*_n(\omega') \braket{\hat A_k^\dagger \hat A_l^\dagger \hat A_m \hat A_n}_\psi\,,
\end{align}
As before we look at the expectation value 
\begin{align}
\braket{\hat A_k^\dagger \hat A_l^\dagger \hat A_m \hat A_n}_\psi = \braket{\text{vac}|\mathcal{\hat U}_{\text{SQ}}^\dagger \hat A_k^\dagger  \mathcal{\hat U}_{\text{SQ}} \mathcal{\hat U}_{\text{SQ}}^\dagger \hat A_l^\dagger  \mathcal{\hat U}_{\text{SQ}} \mathcal{\hat U}_{\text{SQ}}^\dagger\hat A_m  \mathcal{\hat U}_{\text{SQ}} \mathcal{\hat U}_{\text{SQ}}^\dagger \hat A_n \mathcal{\hat U}_{\text{SQ}}|\text{vac}}\,,
\end{align}
and use the linear Bogoliubov transformation and then expand to obtain
\begin{align}
\braket{\hat A_k^\dagger \hat A_l^\dagger \hat A_m \hat A_n}_\psi &= \sinh(r_k) \sinh(r_l) \sinh(r_m) \sinh(r_n)\braket{\text{vac}|\hat B_k \hat B_l \hat B_m^\dagger \hat B_n^\dagger|\text{vac}}\\\label{eq:fo}
&=\sinh^2(r_k) \sinh^2(r_l) \delta_{k,m}\delta_{l,n}+ \sinh^2(r_k) \sinh^2(r_l) \delta_{k,n}\delta_{l,m}\,.
\end{align}
Plugging these results into Eq. (\ref{G2nico}), we find
\begin{align}
\braket{\hat a^\dagger (\omega) \hat a^\dagger (\omega') \hat a(\omega) \hat a(\omega')}_\psi = \braket{\hat a^\dagger (\omega) \hat a(\omega)}_\psi \braket{\hat a^\dagger (\omega')  \hat a(\omega')}_\psi
+ \braket{\hat a^\dagger (\omega) \hat a(\omega')}_\psi \braket{\hat a^\dagger (\omega')  \hat a(\omega)}_\psi\,,
\end{align}
where the terms in the RHS are given by Eq. (\ref{g1nico}).

\section{Derivation of $G^{(1)}(t,t')$ and $G^{(2)}(t,t')$}\label{sec:der}

The  first-order temporal correlation function for mode $a$ is given by \cite{Glauber1963c}
\begin{align}
G_a^{(1)}(t_1,t_2)=\langle\hat{E}_a^{(-)}(t_1)\hat{E}_a^{(+)}(t_2)\rangle_{\psi}\,,
\end{align}
where $\hat{E}_a^{(\pm)}(t_1)$ are the usual positive/negative frequency components of the electric field operator in mode $a$. To simplify calculations, we follow Christ \emph{et al.} \cite{Christ2011} and replace the electric field operators by photon number creation and destruction operators ($\hat{E}_a^{(+)}\propto \hat{a}(t)$). This is valid when the spectra of the beams are not too broad. We thus have
\begin{align}
G_a^{(1)}(t_1,t_2)=\langle\hat{a}\dg(t_1)\hat{a}(t_2)\rangle_{\psi}\,.
\end{align}

We now use the relation
\begin{align}\label{eq:at}
\hat{a}(t)=\frac{1}{\sqrt{2\pi}}\int d\omega \hat{a}(\omega)e^{i\omega t}=\frac{1}{\sqrt{2\pi}}\int d\omega \left(\sum_k \phi_k^*(\omega)\hat{A}_k\right)e^{i\omega t}\equiv\sum_k \hat{A}_k\tilde \phi^*_k(t)\,,
\end{align}
where
$\tilde \phi_k(t)=\frac{1}{\sqrt{2\pi}}\int d\omega \phi_k(\omega)e^{-i\omega t}$. This gives
\begin{align}
G_a^{(1)}(t_1,t_2)={}&\left\langle\left(\sum_k \hat{A}\dg_k\tilde \phi_k(t_1)\right)\left(\sum_l \hat{A}_l\tilde \phi_l^*(t_2)\right)\right\rangle_{\psi}\\
={}&\sum_{k,l}\tilde \phi_k(t_1)\tilde \phi_l^*(t_2)\left\langle \hat{A}\dg_k\hat{A}_l\right\rangle_{\psi}\,.
\end{align}
Using the result in Eq. (\ref{eq:thing}), we have
\begin{align}
G_a^{(1)}(t_1,t_2)={}&\sum_{k,l}\tilde \phi_k(t_1)\tilde \phi_l^*(t_2)\sinh^2(r_k) \delta_{k,l}\\
={}&\sum_k \tilde \phi_k(t_1)\tilde \phi_k^*(t_2)\sinh^2(r_k)\\\label{eq:G1t}
={}&\sum_k \tilde \phi_k(t_1)\tilde \phi_k^*(t_2)\bar n_k\,,
\end{align}
where $\bar n_k$ is defined in Eq. (\ref{eq:nk}). 

The  second-order temporal correlation function for mode $a$ is given by \cite{Glauber1963c} 
\begin{align}
G_a^{(2)}(t_1,t_2)={}&\left\langle\hat{a}\dg(t_1)\hat{a}\dg(t_2)\hat{a}(t_1)\hat{a}(t_2)\right\rangle_{\psi}\,.
\end{align}
Using a similar procedure as above, we can write this as
\begin{align}
G_a^{(2)}(t_1,t_2)={}&\sum_{k,l,m,n}\tilde \phi_k(t_1)\tilde\phi_l(t_2)\tilde \phi_m^*(t_1)\tilde \phi_n^*(t_2)\left\langle \hat{A}\dg_k\hat{A}\dg_l\hat{A}_m\hat{A}_n\right\rangle_{\psi}\,.
\end{align}
Using Eq. (\ref{eq:fo}), we have
\begin{align}
\begin{split}
G_a^{(2)}(t_1,t_2)={}&\sum_{k,l,m,n}\tilde \phi_k(t_1)\tilde\phi_l(t_2)\tilde \phi_m^*(t_1)\tilde \phi_n^*(t_2)\\
&\times\left(\sinh^2(r_k) \sinh^2(r_l) \delta_{k,m}\delta_{l,n}+ \sinh^2(r_k) \sinh^2(r_l) \delta_{k,n}\delta_{l,m}\right)
\end{split}\\
={}&\sum_{k}\tilde \phi_k(t_1)\tilde \phi_k^*(t_1)\bar{n}_k\sum_l\tilde\phi_l(t_2)\tilde \phi_l^*(t_2)\bar{n}_l +\sum_{k}\tilde \phi_k(t_1)\tilde \phi_k^*(t_2)\bar{n}_k\sum_l\tilde\phi_l(t_2)\tilde \phi_l^*(t_1) \bar{n}_l\,,
\end{align}
where $\bar n_k$ is defined in Eq. (\ref{eq:nk}).  Comparing with the result in Eq. (\ref{eq:G1t}), we find that
\begin{align}
G_a^{(2)}(t_1,t_2)={}&G_a^{(1)}(t_1,t_1)G_a^{(1)}(t_2,t_2)+G_a^{(1)}(t_1,t_2)G_a^{(1)}(t_2,t_1)\,.
\end{align}

\end{document}